\documentclass[10pt,a4]{article}

\usepackage{graphicx}

\def\vec#1{{\mathbf #1}}

\def\pfrac#1/#2.{\frac{\partial #1}{\partial #2}}
\def\ne{n_{\rm e}}
\def\np{n_{\rm p}}
\def\nn{n_{\rm n}}
\def\ve{\vec{v}_{\! \rm e}}
\def\vp{\vec{v}_{\! \rm p}}
\def\vn{\vec{v}_{\! \rm n}}
\def\De{D_{\rm e}}
\def\Dp{D_{\rm p}}
\def\Dn{D_{\rm n}}
\def\Sph{S_{\rm ph}}
\def\SeP{S_{\rm e}^{+}}
\def\SnP{S_{\rm n}^{+}}
\def\SpP{S_{\rm p}^{+}}
\def\SeM{S_{\rm e}^{-}}
\def\SnM{S_{\rm n}^{-}}
\def\SpM{S_{\rm p}^{-}}

\def\Diam{{\cal D}}
\def\PRE{Phys. Rev. E}

\newcommand{\JPD}{{\it J. Phys. D: Appl. Phys.} }
\newcommand{\PRL}{{\it Phys. Rev. Lett.} }
\newcommand{\PSST}{{\it Plasma Sources Sci. Technol.} }

\title{Derivation of a merging condition for two interacting streamers in air}

\author{{\bf Zden\v{e}k Bonaventura}\\
Department of Physical Electronics,\\
              Faculty of Science, Masaryk University,\\ 
              Kotl\'a\v{r}sk\'a 2, 611 37 Brno, Czech Republic\\
\texttt{zbona@physics.muni.cz}
\and
{\bf Max Duarte}\\
Laboratoire J.-A. Dieudonn\'e -- UMR CNRS 6621,\\ 
              Universit\'e de Nice -- Sophia Antipolis,\\
              Parc Valrose,
              06108 Nice Cedex 02,
              France
\and
{\bf Anne Bourdon}\\
CNRS, UPR 288 Laboratoire EM2C,\\
              Grande voie des vignes, 92295
              Ch\^{a}tenay-Malabry, France\\
Ecole Centrale Paris,\\
              Grande voie des vignes, 92295
              Ch\^{a}tenay-Malabry, France
\and
{\bf Marc Massot}\\
CNRS, UPR 288 Laboratoire EM2C,\\
              Grande voie des vignes, 92295
              Ch\^{a}tenay-Malabry, France\\
Ecole Centrale Paris,\\
              Grande voie des vignes, 92295
              Ch\^{a}tenay-Malabry, France\\
Center for Turbulence Research, Stanford University,\\
              Building 500, 488 Escondido Mall,
              Stanford CA 94305-3035, USA
}

\begin{document}

\maketitle

\begin{abstract}
The simulation of the interaction of two simultaneously propagating air 
streamers of the same polarity is presented.
A parametric study has been carried out using an accurate numerical method
which ensures a time-space error control of the solution.
For initial separation of both streamers smaller or comparable to
the longest characteristic absorption length of photoionization in air, 
we have found that the streamers tend to merge at the moment when 
the ratio between their characteristic width and their mutual 
distance reaches a value of about 0.35 for positive streamers,
and 0.4 for negative ones.
Moreover it is demonstrated that these ratios are
practically independent of the applied electric field,
the initial seed configuration, and the pressure.
\end{abstract}

Electrical breakdown in air gaps often involves the development of fast ionizing waves
that take the form of thin filaments  called streamers \cite{Raizer}.
These complex and highly nonlinear phenomena precede the formation of sparks and leaders.
Similar filamentary structures are observed in sprites:
large-scale discharges which appear at low pressure in the altitude range
of about $40$ to $90$ km above large thunderstorms \cite{Sentman:1995,Pasko:2007}.
Streamers often form branches. Then  the individual streamer heads
carry charges of the same polarity, and thus they are
electrostatically  repelling each other.
However, careful observations show also the opposite behavior: streamers may even  merge or reconnect \cite{Bries,Cummer,Nijdam}.
So far, only few numerical studies have been carried out on streamer interactions.
First, the impact of charges of numerous positive streamers, propagating simultaneously,
on the electric field and on the velocity of one single positive streamer was studied in
\cite{Naidis} for a 2D configuration.
Later on it was shown in \cite{PRE} that
the electrodynamics of a 2D array of negative streamers substantially 
differs from the one of a single streamer due to their electrostatic
interaction. 
Recently, the interaction of two streamer discharges in air and
other oxygen-nitrogen mixtures has been studied for a 3D cylindrical
configuration in \cite{PRL}.
These authors showed that two competing mechanisms have to be considered: 
attraction due to nonlocal photoionization between streamers and electrostatic 
repulsion of streamer heads due to their space charges of same polarity.
Qualitatively, they concluded that streamer merging is favored if
the distance between streamers is smaller or comparable 
to the longest absorption length of photoionization.
However, a more precise or even quantitative description
of streamer merging remains an open problem.

In this work we carry out a parametric analysis of the dynamics of interacting streamers
to better understand the conditions of streamer merging. 
Therefore we simulate the interaction of two streamers of the same polarity
by means of an accurate numerical method which ensures a time-space error control of the solution.
Taking into account that fully 3D simulations are still 
computationally expensive to carry out 
an extensive parametric analysis, we have considered a Cartesian 2D geometry to reduce computational costs.
This study is then a first step towards a quantitative analysis of streamer merging in real 3D configurations. 
However, in agreement with \cite{PRE,BrauPRE}, we consider that the main 
characteristics of the dynamics of interacting streamers in two dimensions, 
as described below, will be similar in three dimensions. 
In practice, we aim at defining geometrical
parameters independent of the multivariable complex
physical settings
and thus a simple characterization that may be extended
to more general configurations.

We consider the classical fluid model for air
given by drift-diffusion equations
self-consistently coupled with Poisson's equation \cite{Babaeva:1996,Kulikovsky:1997c}:
%%%%%%%%
\def\SeP{\ne\alpha |\vec v_{\rm e}|}
\def\SpP{\ne\alpha |\vec v_{\rm e}|}
\def\SeM{\ne\eta  |\vec v_{\rm e}| -  \ne\np\beta_{\rm ep}}
\def\SpM{\ne\np\beta_{\rm ep} - \nn\np\beta_{\rm np}}
\def\SnM{\nn\np\beta_{\rm np}}
\def\SnP{\ne\eta  |\vec v_{\rm e}|}
\def\SD{\nn\gamma}
%%%%%%%%%
%\begin{widetext}
\begin{equation}\label{trasp} 
\left.
\begin{array}{l}
\partial_t \ne +\nabla\cdot(\ne\,\ve) %
  -\nabla\cdot(\De\ \nabla\ne) = 
         \SeP  -\SeM + \SD + \Sph, \\
\partial_t \np +\nabla\cdot(\np\vp) %
  -\nabla\cdot(\Dp\,\nabla\np) =  
         \SpP - \SpM + \Sph, \\
\partial_t \nn  +\nabla\cdot(\nn\vn) %
  -\nabla\cdot(\Dn\,\nabla\nn)  =  
         \SnP -\SnM - \SD, 
\end{array} 
\right\}
\end{equation}
%\end{widetext}
\begin{equation}
\varepsilon_0\, \nabla \cdot \vec E  = -q_{\rm e}(\np-\nn-\ne), \quad \vec E = - \nabla V,  \label{poisson}
\end{equation}
where
$n_i$ is the  density of charged species $i$ 
(e: electrons, p: positive ions, n: negative ions),
$V$ and $\vec E$
stand, respectively, for the electric potential and field,
and $\vec{v}_i= \mu_i \vec E$ is the drift velocity.
We denote by
$D_i$ and $\mu_i$ the
diffusion coefficient and the mobility of charged species $i$,
$q_{\rm e}$ is the absolute value of the electron charge,
and   $\varepsilon_0$  is the permittivity of free space.
Moreover
$\alpha$ is the impact ionization coefficient, $\eta$
stands for  the electron attachment coefficient, $\beta_{\rm ep}$
and $\beta_{\rm np}$
are, respectively, the electron-positive ion
and
negative-positive ion recombination coefficients,
and $\gamma$ is the detachment coefficient.
All these coefficients depend
on the local reduced electric field $E/N$,
and thus vary in time and space,
where $E=|\vec E|$ is the electric field magnitude, and $N$ is the
air neutral density.
Further details are given in \cite{JCP}.
For $\Sph$, the photoionization source term for air, 
we use the 3-Group SP$_3$ model derived in \cite{Bourdon:2007} with Larsen's
boundary conditions  \cite{Liu:2007}.
In this model, three equivalent absorption lengths are used to model 
the wavelength dependence of the photoionization source term in air.
At ground pressure these absorption lengths are 
$(\lambda_1 p_{{\rm O}_2})^{-1}=0.1408\,$cm,
$(\lambda_2 p_{{\rm O}_2})^{-1} = 0.0561\,$cm,
and $(\lambda_3 p_{{\rm O}_2})^{-1}=0.0105\,$cm,
where $p_{{\rm O}_2}  = 158.9\,$Torr is the partial pressure of
molecular oxygen in air.

To handle the stiff modeling equations,
we have recently developed in \cite{JCP} a
numerical scheme for multi-scale streamer discharge  simulations for
general Cartesian multi-dimensional geometries.
The strategy is based on a second
order time adaptive integration with a splitting technique
and dedicated solvers,
and space finite volume multiresolution
for dynamic grid adaptation.
These features involve important efficiency gains
in terms of CPU time and
memory space while ensuring a time-space error
control of the solution.
Poisson's equation and Helmholtz
equations issued by the photoionization model
are considered on the adapted grid,
and the resulting sparse linear systems are
solved with MUMPS \cite{mumps1, mumps2}.
In all simulations carried out in this work (except for some computations in figure  \ref{fig-y0} where a finer grid is used)
a time-space accuracy tolerance of $10^{-4}$ is used for a
space resolution of 3.9$\,\mu$m.
This choice guarantees a sufficiently fine time-space
representation of the physics,
and numerical results disclosing practically the same
physical behavior for higher spatial resolutions and tighter
accuracy tolerances.
The domain size is carefully chosen for the various
configurations and with the same space resolution of 3.9$\,\mu$m
such that
no interference of the boundaries is evidenced.

In order to preserve similarity constraints at different
pressures (corresponding to discharges at different
altitudes),
all variables are scaled by an appropriate
power of the ratio of the  air density at ground pressure $N_0$,
and the air density $N$  at the  given pressure  \cite{Liu:2006}.
For example, the space dimensions scale as  $\vec x = {\vec x}_0 N_0/N$, the charge densities
as $n = n_{0} N^2/N_0^2$, the electric field  as $E=E_0 N/N_0$,
and  the time as $ t = t_0 N_0/N$, where the subscript ``0'' refers to
the values at ground pressure.  Our reference (ground) density for air is
$N_0=2.688\times 10^{19}\,\hbox{cm}^{-3}$ at a temperature of $273\,$K.

We study two separate discharge filaments initiated by placing two
identical Gaussian plasma clouds into a homogeneous background
electric field generated by two remote electrodes.
The pressure, streamer polarity, applied electric field, and the configuration  parameters
of the initial Gaussian seeds (the width $\sigma$, the maximum density $n_{\rm max}$, and the mutual
separation $y_0$) are varied in order to derive a condition for the merging of both streamers.
\begin{figure}
\includegraphics[width=0.8\textwidth]{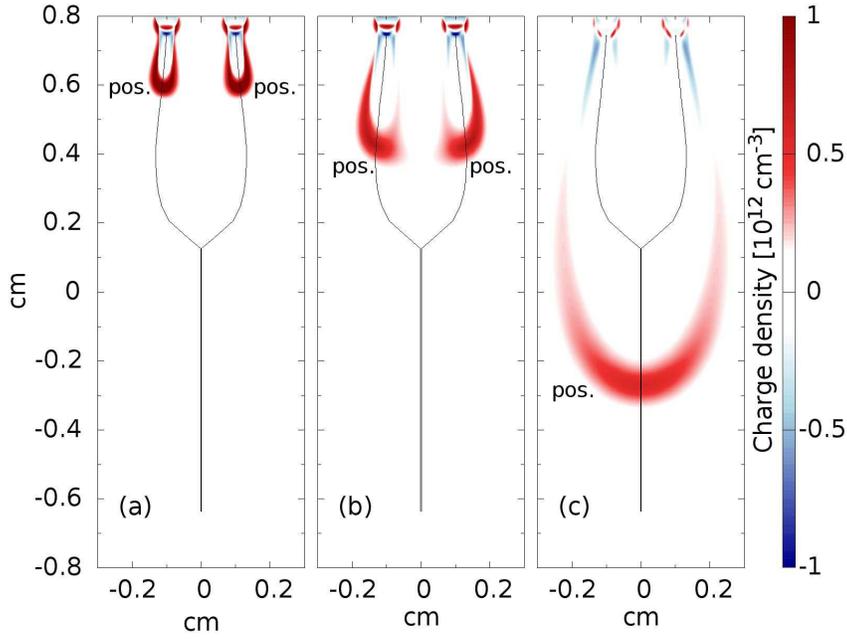}
\caption{(color online) Time evolution of the net charge density for
two positive streamers at ground pressure ($N/N_0=1.0$) for an applied electric field  $E_{\rm a}=1.5E_{\rm bd}$ and two Gaussian seeds with
$y_0 = 0.2\,N_0/N\,$cm,
$n_{\rm max}=10^{13}\,N^2/N_0^2\,$cm$^{-3}$, and $\sigma = 0.02 N_0/N\,$cm. (a)
 $t=5.0\,$ns: well-developed streamers repulsing each other,
(b) $t=t_{\rm tr}=6.3\,$ ns: transition between repulsion and merging, (c)
$t=8.0\,$ns: propagation of a single discharge.
Black solid line: trajectory of the maximum electric field.
\label{fig1}}
\end{figure}

\begin{figure}
\includegraphics[width=0.8\textwidth]{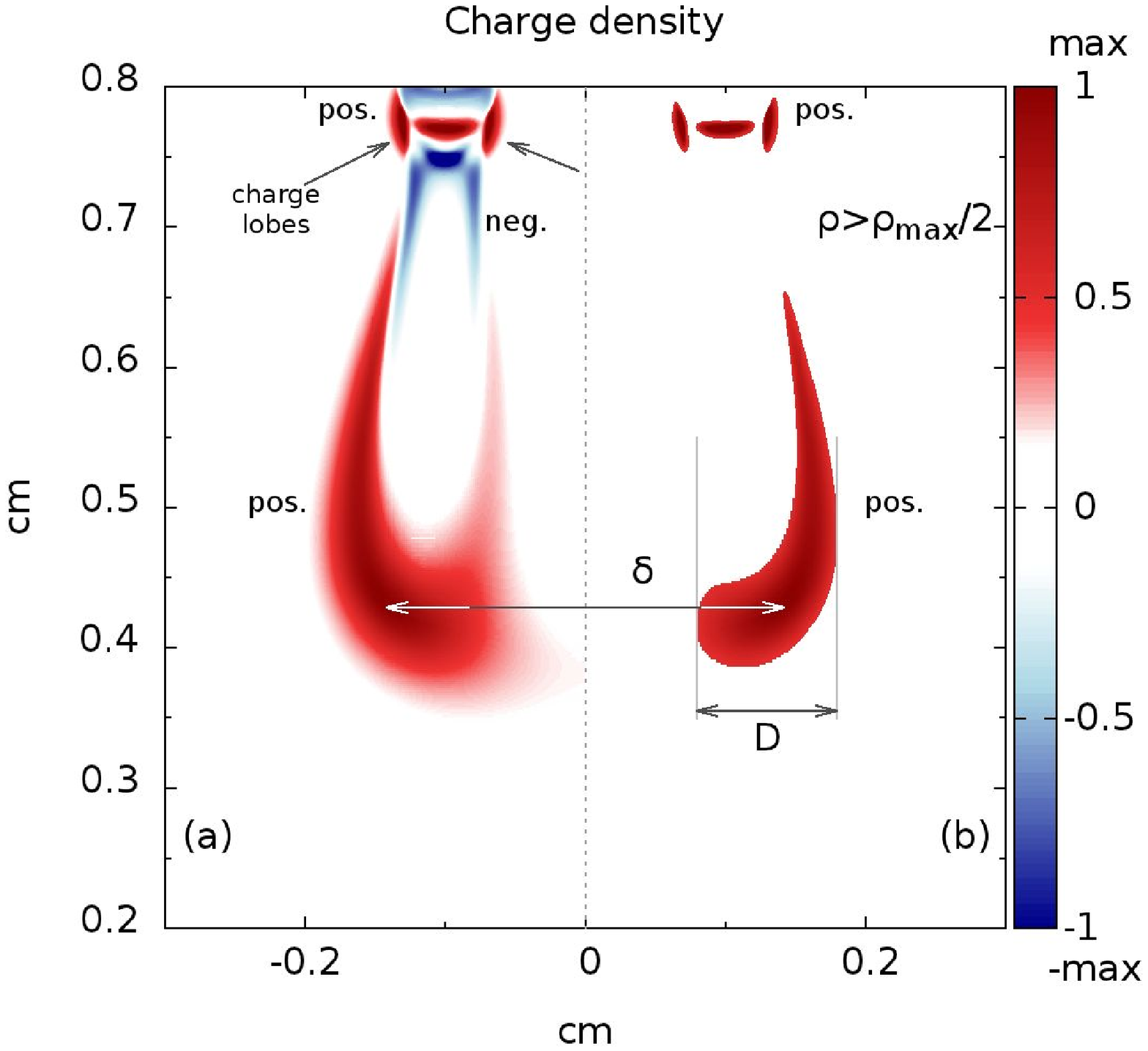}

\caption{(color online)  (a) Net charge density  and 
   (b) Region where the net charge density is  higher than
       half the maximum net charge density at $t=t_{\rm tr}$ 
       for the same condition as in Figure~\ref{fig1}(b). 
       Definition of the {\it streamer head characteristic width\/} $\Diam$ and 
       the {\it mutual distance of both filaments\/} $\delta$.
       \label{fig2}}
\end{figure}

Figures~\ref{fig1}(a)--(c) show snapshots of different phases of the evolution of the net charge
density
($\rho =\np  -\ne -\nn$)
for two simultaneously propagating positive discharge filaments at ground pressure  ($N/N_0=1.0$)
for an initial
seed separation of $y_0 = 0.2\,N_0/N\,$cm,
a maximum density of  $n_{\rm max}=10^{13}\,N^2/N_0^2\,$cm$^{-3}$,
a width $\sigma = 0.02 N_0/N\,$cm,
and an applied electric field  $E_{\rm a}=1.5E_{\rm bd}$, where $E_{\rm bd}=32\, N/N_0\,{\rm kV\,cm^{-1}}$
is the breakdown electric field in air.
The complete trajectory
of the maximum electric field is also shown.
Following this trajectory, the transition time $t_{\rm tr}$,
between the streamer repulsion
and the beginning of their merging,
is defined as the instant when the separation of the
two trajectories starts to decrease.
For the condition of Figure~\ref{fig1}, we have $t_{\rm tr}=6.3\,$ns.
Figure~\ref{fig2}(a) shows at this time the distribution of the net charge density.
We define in Figure~\ref{fig2}(b) the {\it mutual distance of both filaments\/} $\delta$
 by the separation of the maximum net density $\rho_{\rm max}$
in each head; and
the {\it streamer head characteristic width} $\Diam$
as the planar length set by the cross section in the direction perpendicular 
to the applied electric field  ${\vec E}_{\rm a}$,
where the charge density is locally
higher than half the maximum net density:
$\rho \geq \rho_{\rm max}/2$.
%%%%%%%%%%%%

\begin{figure}
\includegraphics[width=0.8\textwidth]{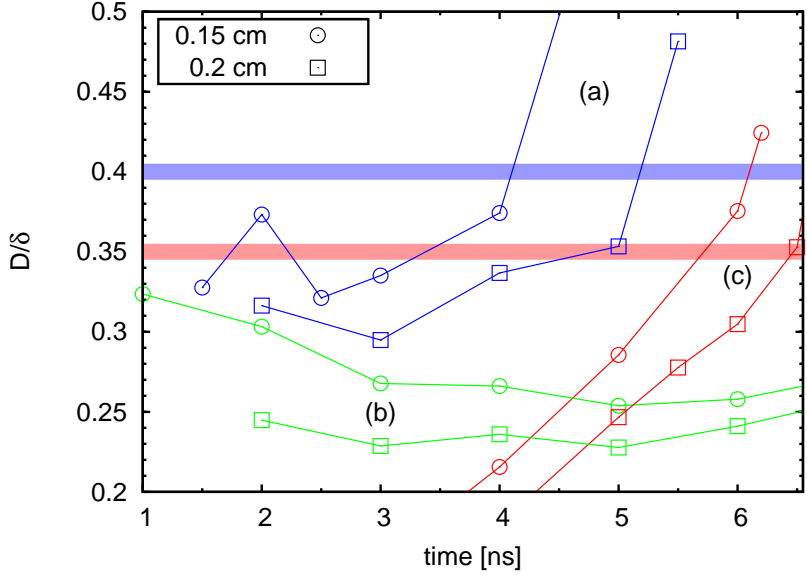}
\caption{ (color online)
Time evolution of the ratio $\Diam/\delta$ for (a) two negative streamers with photoionization source term, (b) two negative
      streamers without photoionization source term,
      (c) two positive streamers with photoionization source term. Results for 
      initial separation of the seeds $y_0 = 0.15\,N_0/N\,$cm and $y_0 = 0.2\,N_0/N\,$cm are shown.
      Other parameters are similar
      to Figure 1. Critical merging ratios are indicated with stripes.  \label{fig-dyn}}
\end{figure}

\begin{figure}
\includegraphics[width=0.8\textwidth]{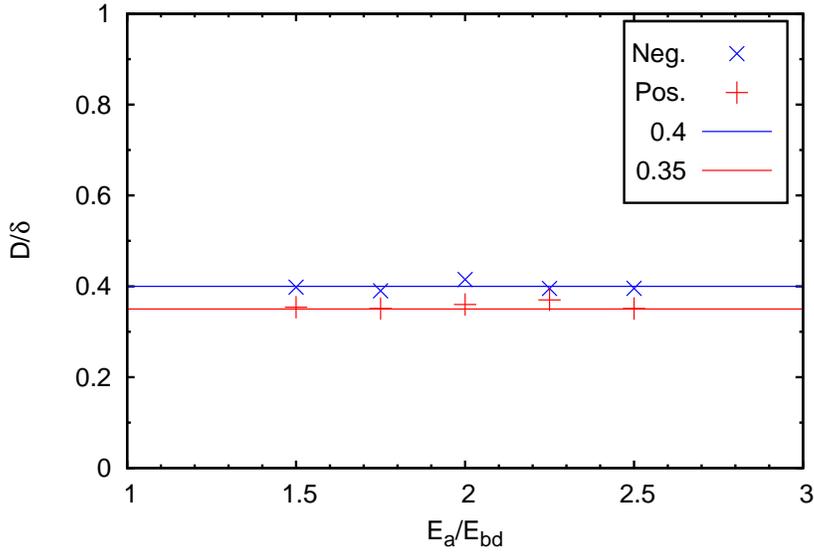}
\caption{(color online)
  Ratios $\Diam/\delta$ for positive and negative streamers for
  different values of the applied electric field $E_{\rm a}/E_{\rm bd}$. Other parameters are similar to Figure~\ref{fig1}.
  \label{figE}}
\end{figure}

\begin{figure}
\includegraphics[width=0.8\textwidth]{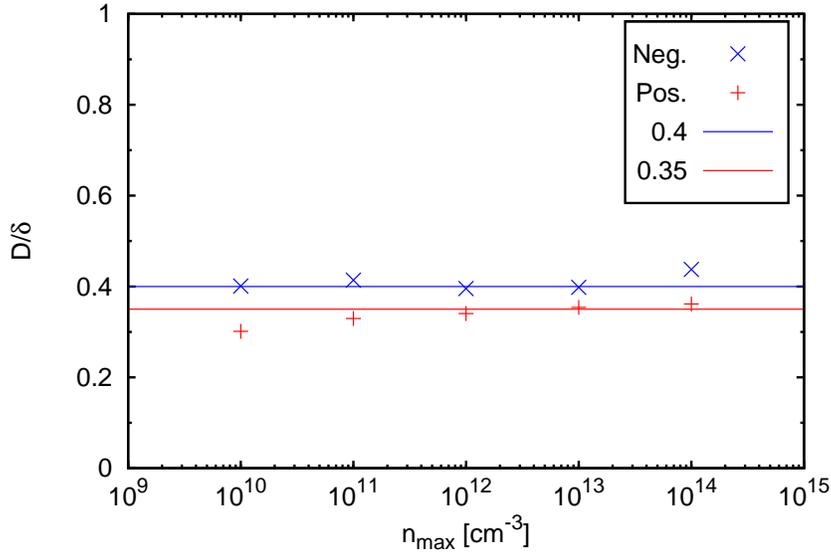}%
\caption{(color online)
   Ratios $\Diam/\delta$ for  positive and negative streamers for different
   values of the maximum density $n_{\rm max}$ of seeds. Other parameters are similar to Figure~\ref{fig1}.
  \label{fig-n0} }
\end{figure}

\begin{figure}
\includegraphics[width=0.8\textwidth]{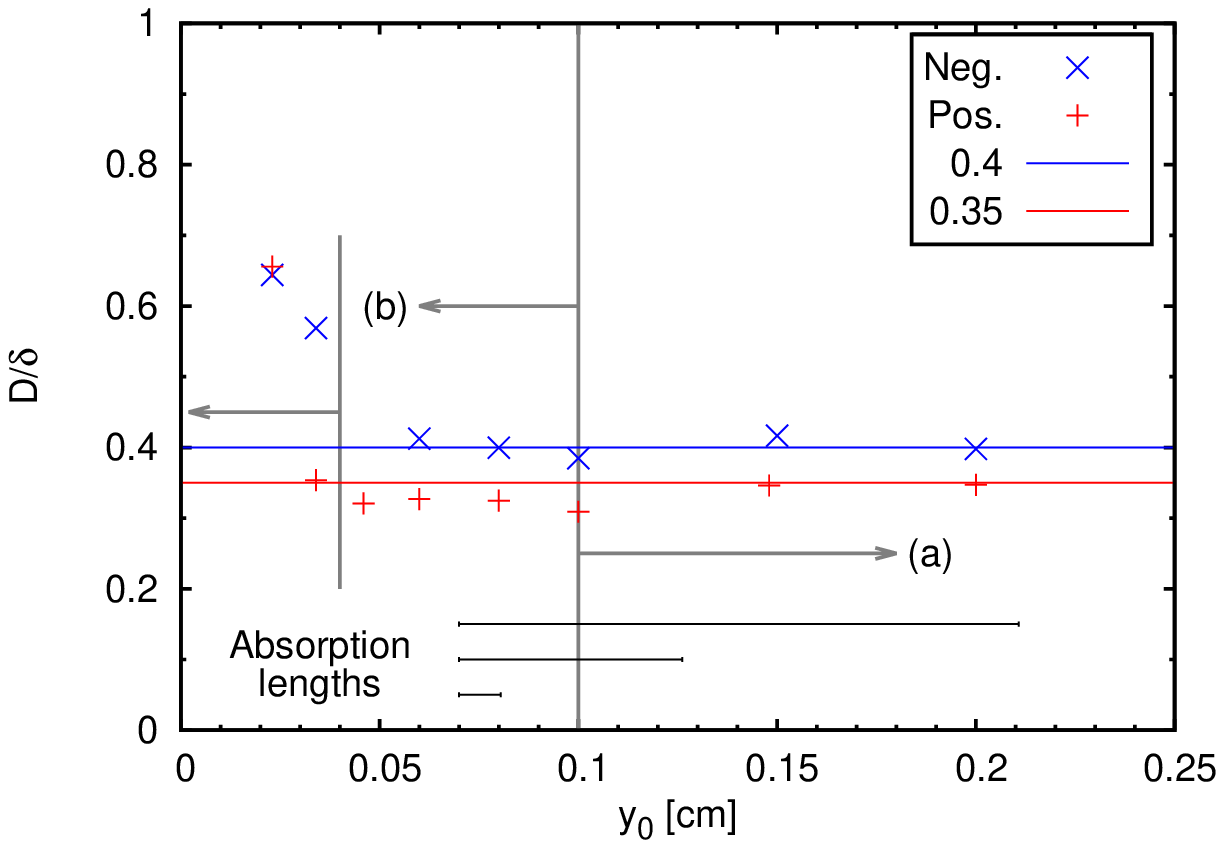}
\caption{(color online)
  Ratios $\Diam/\delta$ for  positive and negative streamers for different
  initial seed distances $y_0$, with (a) $n_{\rm max}=10^{13}\,N^2/N_0^2\,$cm, $E=1.5E_{\rm bd}$,
  $\sigma = 0.02\, N_0/N\,$cm;
  (b)  $n_{\rm max}=10^{10}\,N^2/N_0^2\,$cm, $E=2.5E_{\rm bd}$, $\sigma = 0.00736\, N_0/N\,$cm.
  Absorption  lengths for photoionization at ground pressure are shown.
 \label{fig-y0} }
\end{figure}

We have found that the merging condition for two streamers in air could 
be represented by the value of the ratio $\Diam$/$\delta$ which appears 
to be very stable for a given polarity of discharges.
To illustrate this, Figure~\ref{fig-dyn} shows the time evolution of the 
ratio $\Diam/\delta$ for two positive or two negative streamers.
In this figure, initial separation of the seeds  $y_0 = 0.15\,N_0/N\,$cm and $0.2\,N_0/N\,$cm
are considered,  and other parameters are similar to Figure 1.
First, Figure~\ref{fig-dyn} shows the time evolution of the ratio $\Diam/\delta$ 
for two negative streamers as these discharges may propagate without photoionization source term.
When photoionization is included,
the ratio $\Diam/\delta$ increases  
and reaches a value of 0.4 at the moment of transition
($4\,$ns in case of $y_0 = 0.15\,N_0/N\,$cm, and  $5\,$ns  in case of $y_0 = 0.20\,N_0/N\,$cm)
between streamer repulsion 
and the beginning of their merging, and streamer merging is then observed. 
It is interesting to note that when photoionization is omitted, the 
ratio $\Diam/\delta$ is less than 0.4, and no merging is evidenced.
Conversely
photoionization must be included 
for positive streamers
as it is necessary for their propagation. In this case 
Figure~\ref{fig-dyn} shows that  the ratio $\Diam/\delta$ increases 
and reaches a value of 0.35 at the transition time 
$5.6\,$ns for $y_0 = 0.15\,N_0/N\,$cm, ($6.3\,$ns for $y_0 = 0.20\,N_0/N\,$cm),
and then streamer merging is observed.
These results clearly show the crucial role of the photoionization source term on streamer merging.
In the following we show that the value of the ratio $\Diam/\delta$ at the transition 
time depends on the polarity of discharges but is practically independent 
on the multivariable physical settings.
For instance, for the same parameters of the seeds as in Figure~\ref{fig1}
($n_{\rm max}=10^{13}\,N^2/N_0^2\,$cm$^{-3}$, $y_0 = 0.2\,N_0/N\,$cm, and $\sigma = 0.02\, N_0/N\,$cm)
we have performed a
set of simulations with an applied electric field $E_{\rm a}$
varying from $1.5E_{\rm bd}$ to $2.5E_{\rm bd}$
at ground pressure.
Figure~\ref{figE} shows that the resulting values of $\Diam/\delta$ at the transition times for merging streamers are roughly constant and
of about $0.35$ for positive polarity discharges,
and approximatively $0.4$ for negative ones.
Hence $\Diam/\delta$ is independent of the applied electric field $E_{\rm a}$ for both streamer polarities.
Similarly,
Figure~\ref{fig-n0} shows the dependence of $\Diam/\delta$ on $n_{\rm max}$, the maximum density of seeds, for a fixed applied electric field $E_{\rm a}=1.5E_{\rm bd}$.
Same previous ratios are obtained
whereas
$\Diam/\delta$ is slightly decreasing as $n_{\rm max}$ decreases
for positive streamers.
However
a variation of four orders of magnitude for $n_{\rm max}$ involves
only a variation of about $15\%$ of $\Diam/\delta$.
In the same way we have tested the variation of $\Diam/\delta$ with altitude (not shown here).
We have considered altitudes ranging from the ground up to 80 km, corresponding to pressure
range from 1 to $1.5\times 10^{-5}$ atm.
We have found that the ratio $\Diam/\delta$  is independent of the altitude and then of pressure
in the range studied.
%%%%%%%%%%%%%%%%%%%%%%%
%%%%%%%%%%%%%%%%%%%%%%%

%%%%%%%%%%%%%%%%%%%%%%%%%%%%%%%%%%%%%%%%%%

%%%%%%%%%%%%%%%%%

%%%%%%%%%%%%%%%

%%%%%%%%%%%%%%%%%%%%%%%%%%

Moreover we have performed simulations to study the influence of the initial 
seed separation $y_0$ on $\Diam/\delta$.
Figure~\ref{fig-y0} shows the dependence of $\Diam/\delta$ for 
$y_0\leq 0.2 \,N_0/N\,$cm. Two sets of seed parameters were considered:
(a) for  $y_0 > 0.1\,N_0/N\,$cm: $n_{\rm max}=10^{13}\,N^2/N_0^2\,$cm, 
    $\sigma = 0.02\,N_0/N\,$cm, and $E=1.5 E_{\rm bd}$;
(b) for $y_0 < 0.1\,N_0/N\,$cm: $n_{\rm max}=10^{10}\,N^2/N_0^2\,$cm, 
    $\sigma = 0.00736\, N_0/N\,$cm, and $E=2.5E_{\rm bd}$.
The first set corresponds to the reference condition used  for Figure~\ref{fig1}.
The second set was chosen to be close to the conditions used in \cite{PRL},
and is characterized by a high value of the applied electric field, comparable
to typical values of the electric field in streamer heads,
and small initial seed width and separation. Because of the small seed
width of the second set, for this set, computations were performed
for a space resolution  of 1.95$\,\mu$m.
Figure~\ref{fig-y0} shows that for positive streamers with $y_0 \geq 0.04\, N_0/N\,$cm,
the value of the ratio $\Diam/\delta$ for the second set is slightly 
less than the reference value of $0.35$ for the first set,
which can be explained by the small dependence of $\Diam/\delta$ on $n_{\rm max}$,
as seen in Figure~\ref{fig-n0}.
For negative streamers, Figure~\ref{fig-y0} shows that for
$y_0 \geq 0.04\, N_0/N\,$cm, the ratio $\Diam/\delta$ is independent of the value of $y_0$.
For initial separation of streamers smaller than
$y_0 < 0.04\, N_0/N\,$cm
the transition between repulsion and merging is occurring
very close to the initial Gaussian seeds and therefore the region
where  $\rho \geq \rho_{\rm max}/2$ is indistinguishably joined
with  the charge lobes, typically observed near the initial germs
(shown in Figure~\ref{fig2}).
In this case, the streamer head appears to be significantly wider,
and consequently the ratio $\Diam/\delta$ increases as shown in Figure~\ref{fig-y0}.
It is interesting to note that in the two sets used for Figure~\ref{fig-y0}, very different values  of $\sigma$ have been used.
We have carried out additional simulations to verify that the value of $\sigma$ has a negligible influence on the ratio $\Diam/\delta$.

Finally, for simulations presented in this work, we have found that a stable value for the ratio $\Diam/\delta$ is obtained 
for merging streamers for an initial separation $y_0$ which is smaller or comparable
to the longest absorption length of photoionization for air
($0.1408 N_0/N\,$cm). These results clearly show the significant role of photoionization on streamer merging. We have carried out simulations for larger values of $y_0$, but it was not possible to derive simple conclusions on streamer interactions from the analysis of the value of the ratio $\Diam/\delta$.

In conclusion, in this work, we have carried out a parametric study on the interaction of two streamers in air
using 2D numerical simulations based on an accurate numerical method which ensures a time-space error control of the solution. 
This study is a first step towards a quantitative analysis of streamer merging in real 3D conditions. 
We have shown that  for a initial separation between streamers that is 
smaller or comparable to the longest characteristic absorption length of photoionization
($0.1408 N_0/N\,$cm), streamers will start to merge, i.e., the distance between their trajectories
starts to decrease despite their electrostatic repulsion, at the moment when the ratio $\Diam/\delta$
of  the  streamer characteristic width $\Diam$ and  the streamer mutual distance $\delta$
attains a value of about 0.35 for positive streamers, and 0.4 for negative ones.
These ratios are independent of
the applied field, the initial seed configuration, and pressure.
Moreover presented results clearly illustrate the significant role of photoionization on the streamer merging.
Further studies are required to find the key parameters to characterize streamer interactions
when $y_0$ is considerably larger than the longest photoionization absorption length.

\subsection*{Acknowledgements} 
This research was  supported by a fundamental project grant from ANR (French National Research Agency):
{\it S\'echelles} (ANR-09-BLAN-0075-01, PI. S. Descombes), and by a DIGITEO RTRA project: {\it MUSE} (PI. M. M.). Z. B.
acknowledges support by project CZ.1.05/2.1.00/03.0086 funded by European Regional Development Fund and support of Ecole
Centrale Paris.


\begin{thebibliography}{18}

\bibitem{Raizer} Raizer Y P 1991  \textit{Gas Discharge Physics}  (Springer-Verlag, Berlin Heidelberg, 1991)

\bibitem{Sentman:1995}
  Sentman D D,   Wescott E M,  Osborne D L,  Hampton D L, and  Heavner M J 1995
  \textit{Geophys. Res. Lett.} \textbf{22} 1205--8

\bibitem{Pasko:2007}  Pasko V P 2007 \textit{Plasma Sources Sci. Technol.} \textbf{16}  S13--S29

\bibitem{Bries}  Briels  T M P, Kos J,  van Veldhuizen E M and Ebert U 2006
                   \textit{\JPD} \textbf{39}  5201--10

\bibitem{Cummer} Cummer S A, Jaugey  N, Li J,  Lyons W A,  Nelson T E  and  Gerken E A
                 2006  \textit{ Geophys. Res. Lett.} \textbf{33} L04104

\bibitem{Nijdam} Nijdam S,  Geurts C G C,  van Veldhuizen E M  and  Ebert U 2009
                \textit{\JPD} \textbf{42} 045201

\bibitem{Naidis} Naidis G V 1996 \textit{\JPD} \textbf{29}  779--83

\bibitem{PRE} Luque A,   Brau F and Ebert U 2008  \textit{\PRE} \textbf{78}  016206

\bibitem{PRL} Luque A, Ebert U and Hundsdorfer W 2008  \textit{\PRL} \textbf{101} 075005

\bibitem{BrauPRE}  Brau F,  Luque A, Meulenbroek B,  Ebert U,  and  Sch\"afer L 2008
                  \textit{\PRE}  \textbf{77} 026219

\bibitem{Babaeva:1996} Babaeva N Y and Naidis G V 1996  \textit{\JPD} \textbf{29} 2423--31

\bibitem{Kulikovsky:1997c} Kulikovsky A A 1998 \textit{\PRE} \textbf{57} 7066--74

\bibitem{Bourdon:2007}  Bourdon A, Pasko V P, Liu N Y, Celestin S, Segur P and Marode E 2007
              \textit{\PSST} \textbf{16} 3 656--78

\bibitem{Liu:2007} Liu N Y,  Celestin S, Bourdon A, Pasko V P, Segur P and Marode E 2007 \textit{Appl. Phys. Lett.\/}
                   \textbf{91} 21  211501

\bibitem{JCP} Duarte M, Bonaventura Z, Massot M,  Bourdon A,  Descombes S and Dumont T 2012
              \textit{J. Comput. Phys.\/} \textbf{231} 3 1002--19

\bibitem{mumps1} Amestoy P R, Duff I S, Koster J and  L'Excellent J-Y 2001
   \textit{SIAM J. Matrix Anal. Appl.\/} \textbf{23} 1  15--41

\bibitem{mumps2} Amestoy P R, Guermouche A, L'Excellent J-Y and  Pralet S 2006
                  \textit{Parallel Computing} \textbf{32} 2  136--56

\bibitem{Liu:2006} Liu N and Pasko V P 2006  \textit{\JPD} \textbf{39} 327--34


\end{thebibliography}
\end{document}